\documentclass[a4paper,aps,prd,twocolumn,groupedaddress,showpacs,nofootinbib,showkeys,amssymb]{revtex4}

\usepackage{graphicx}
\usepackage{bm,natbib}
\usepackage{amsmath}
\usepackage[latin1]{inputenc}
\usepackage[english]{babel}
\usepackage[T1]{fontenc}
\usepackage{amssymb}
\usepackage{amsfonts}
\usepackage{epsfig}
\usepackage{colordvi}
\usepackage{psfrag}
\usepackage{color}
\usepackage{dcolumn}
\usepackage{multirow}
\usepackage{hyperref}
\usepackage{epstopdf}
\usepackage{bm}
\usepackage{times}
\hypersetup{
  colorlinks=true,        
  linkcolor=blue,         
  citecolor=magenta,      
}

\begin{document}

\title{Noether's stars in $f(\cal {R})$ gravity}

\author{Mariafelicia De Laurentis\email{mfdelaurentis@tspu.edu.ru}}
\affiliation{Institute for Theoretical Physics, Goethe University, Max-von-Laue-Str. 1, 60438 Frankfurt, Germany}
\affiliation{Dipartimento di Fisica "E. Pancini", Universit\'a di Napoli ?Federico II?, Compl. Univ. di Monte S. Angelo, Edificio G, Via Cinthia, I-80126, Napoli, Italy}
\affiliation{Lab.Theor.Cosmology,Tomsk State University of Control Systems and Radioelectronics (TUSUR), 634050 Tomsk, Russia}
\affiliation{INFN Sezione  di Napoli, Compl. Univ. di
Monte S. Angelo, Edificio G, Via Cinthia, I-80126, Napoli, Italy.}

\begin{abstract}

The Noether Symmetry Approach can be used to construct spherically symmetric solutions in $f({\cal R})$ gravity.
Specifically, the Noether conserved quantity is related to the gravitational mass and a gravitational radius that reduces to  the Schwarzschild radius in the limit $f({\cal R})\rightarrow {\cal R}$.
We show that it is possible to construct the $M-R$ relation for neutron stars depending on the Noether conserved quantity and the associated gravitational radius. 
This approach enables the recovery of extreme massive stars that could not be stable in the standard Tolman-Oppenheimer-Volkoff based on General Relativity. Examples are given for some power law $f({\cal R})$ gravity models.
\end{abstract}


\date{\today}

\maketitle
\section{Introduction}
\label{I}
Compact stars are natural laboratories to  test strong gravity effects or, in general, alternative  theories of gravity.
In particular, some neutron stars present 
 properties,  as the Mass-Radius  ($M-R$) relation,  that  can be hardly explained in the  context of General Relativity adopting simple equations of state. For examples,   PSR J$0348 + 0432$  \cite{1} and 
PSR J$1614-2230$  \cite{2}  represent  a  challenge  for standard theory   and could be a  possible testbed for modified gravity
\cite{PhysRepnostro,OdintsovPR,6,Mauro,faraoni,10,libroSV,libroSF,Sotiriou2010,Telereview}. 
On the other hand,  understanding  the structure of neutron stars  allows  to constrain
the  parameters of any given gravitational theory in  the strong field regime \cite{asta,fiziev,ottewill,stergio}.

However, the most important problem in this research concerns the choice of  equation of state for matter, that, up to now, are not  known with certainty. In  order to explain observations, one can either ask for exotic (unknown)  equations of state or for modifying gravity in the  strong field regime inside  the star  \cite{ruben,idro,jeans}.
To constrain the observational parameters  in modified theories of gravity,  one can use the $ M-R $ relation  as discussed in \cite{MRnostro}.
A drawback in the study of neutron stars models is the fact that one cannot always perform self-consistent matching of  internal and external solutions.  
This is because, in modified gravity, the exterior space-time geometry is not described exclusively by the mass of the star.

This point needs to be clarified. According to the stellar structure, if a theory of   gravity  is viable and can describe, for example, a  neutron star,  a unique solution should be achieved and  internal and external solutions should be consistently matched.   This fact strictly depends on the well formulation and the well position of the Cauchy problem. In a modified theory of gravity,   assigning the mass $M$ and the radius $R$ could not be sufficient to obtain self-consistent boundary conditions. The problem gets worse if the field equations are higher than second order in derivatives because one needs  initial data up to $(n-1)$ order, being $n$ the derivative order of the field equations\footnote{In the case of $f(\cal R)$ gravity, being the field equations of order 4, we need initial data up to the third derivative.}.  This means that it could result extremely difficult to get a unique solution matching internal and external ones. This lack of effective mathematical tools to achieve unique solutions can be partially circumvented considering in detail the Cauchy problem. As discussed in \cite{libroSV,vignolo}, a  choice of source fluid and suitable coordinates in the gravitational field equations can lead to a well position and well formulation of the problem. However, a general recipe, working for any modified theory of gravity, does not exist at the moment.

Furthermore, the  Birkhoff Theorem  \cite{birk} is not always valid in  modified gravity and the consistency of solutions must be carefully verified according to the boundary conditions \cite{vignolo}.
This means that other information concerning the mass distribution is necessary in order to obtain a unique solution for both the interior and exterior regions of stars.

In general, the external solution is imposed by hand to be coincident with the
internal Schwarzschild or Tolman-Oppenheimer-Volkoff (TOV) solution: the method is equivalent to freezing-out the further
degrees of freedom emerging from Modified Gravity with respect to those of General Relativity outside the star.
This approach is controversial because it means that the
full field equations are not considered, and hence the self-consistency of the whole problem is strongly violated.
Consequently, artificial effects on the structure of the star can arise.
A self-consistent analysis of compact objects, in particular of neutron stars and their properties in Modified Theories, in
particular in $f({\cal R})$ gravity\footnote{To avoid confusion between the radius $R$ of the star and the Ricci scalar curvature ${\cal R}$, we adopt a different notation.}, is a fundamental challenge which needs to be addressed.

It is worth stressing that modified theories of gravity were introduced to explain the accelerated expansion of the Universe, the presence of dark matter and, finally, the impossibility to renormalize gravity \cite{PhysRepnostro,OdintsovPR,6,Mauro,faraoni,10,libroSV,libroSF,Sotiriou2010,Telereview}. All the fundamental interactions  have already been described at fundamental level by  quantum field theory, except  gravity. In other words, a self-consistent theory of quantum gravity is not at hand until now.  This means that General Relativity   is not the final theory of gravitation, but only an  approximation of it working very well at local and infrared scales. The simplest generalization  of General Relativity is assuming that  the Hilbert - Einstein action of gravity, linear in the Ricci curvature scalar $\cal R$,  can be generalized as  $f(\cal R)$ where $f$ is an analytic function of $R$ not necessarily linear.  The fundamental reason for this approach lies on the fact that the  formulation of quantum field theory on curved space-times gives rise to higher order corrections to the gravitational action like ${\cal R}+\alpha {\cal R}^2$ \cite{PhysRepnostro}. Furthermore, the effective action of any unified theory, involving gravity,  implies corrections to the Hilbert-Einstein Lagrangian, then $f(\cal R)$ gravity is a natural approach to be pursued.   On the other hand, the form of $f({\cal R})$ can be  constrained assuming a sort of "inverse scattering procedure" considering fine experiments and observations  that can fix the parameters of gravitational interaction \cite{scholar}.  It is interesting to see that a wide range of astrophysical phenomena can be addressed by $f({\cal R})$ gravity  ranging from Solar System scales up to cosmological scales without assuming the dark energy and dark matter hypotheses \cite{annalen}. 
The investigation predicts the existence of new stable neutron star branches with respect to General Relativity \cite{asta}. 
In particular, techniques related to the existence of symmetries and conserved quantities can aid in  the construction of self-consistent 
neutron star models.
The so-called Noether Symmetry Approach \cite{cimento} is one these techniques suitable for these purposes.

In fact, identifying Noether symmetries enables one to "reduce" dynamics by finding out first integrals and, if a complete set of first integrals is identified, to  solve this one through a suitable changes of variables. In other words, if the number of conserved quantities coincides with the dimension of the configuration space, the resulting system is fully integrable.  On the other hand, such conserved quantities are always  related to the physical parameters of dynamical systems. 
In general, the technique has been successfully applied to dark energy and inflationary cosmology \cite{cimento,quinte} and to  dynamical systems in spherical and axial symmetry \cite{arturo}. 

In this paper, the Noether Symmetry Approach is adopted
to fix the radius $R$ and the mass $M$ of neutron stars. 
As it can be shown, both quantities can be related to the Noether conserved quantity emerging in $f({\cal R})$ gravity. 
In this case we say that we are in the presence of a {\it Noether Star}.
Specifically, because the existence of a Noether symmetry is related to the identification
of a vector field in the configuration space whose Lie derivative is conserved, it is possible to perform a change of variables where one (or more than one) cyclic variable appears in the dynamics.  A conserved quantity is related to this variable and then a first integral is derived. 
We will show that such a conserved quantity coincides with the gravitational mass and therefore the gravitational radius of the stellar system. 
In particular, the Noether vector allows to fix a power-law form $f({\cal R})=f_0{\cal R}^{1+\epsilon}$, where the deviations with respect to General Relativity can be easily identified. 
The mass and the radius of the system are   functions of $\epsilon$. The standard Schwarzschild radius and mass of General Relativity  are  recovered for $\epsilon \rightarrow 0$. 
A power law  Lagrangian, like that we are using here,
has been largely tested at different scales. Several works have been done on the study of deviations on the apsidal motion
of eccentric eclipsing binary systems \cite{leome}, as well as tests on the geodesic motions of a massive particles \cite{clifton}. Primordial gravitational waves in the early universe  have been widely studied \cite{maurome}. As discussed in \cite{quinte},  power-law $f(\cal R)$ models have several application in cosmology and can partially alleviate the problem of today observed accelerated expansion also if they have to be improved in order to address the whole cosmic evolution (see \cite{OdintsovPR, Mauro} for details).

The outline of the paper is as follows. 
In Sec.~\ref{II}, the field equations for $f({\cal R})$ gravity are derived.
Sec.~\ref{III} is devoted to the Noether Symmetry Approach. 
The power-law form of $f({\cal R})$, associated conserved quantities and the spherically symmetric solutions are derived. 
The modified TOV solution related to $f({\cal R})=f_0{\cal R}^{1+\epsilon}$ is discussed in Sec.~\ref{IV}. 
Herein the $M-R$ diagram, considering values of $\epsilon \neq 0$ and then demonstrating the deviation of the diagram with respect to General Relativity case ($\epsilon=0$),  is also discussed.
The conclusions are drawn in Sec.~\ref{V}.
\section{Field equations and spherical symmetry in $f(\cal R)$ gravity}
\label{II}
Let us start from  the following action
\begin{equation}\label{action}
\mathcal{A}=\frac{1}{16\pi}\int d^4x\sqrt{-g}\left[f({\cal R})+{\cal L}_m\right],
\end{equation}
where $g$ is the determinant of the metric tensor
and ${\cal L}_m $ is the standard fluid matter Lagrangian. We adopt for the moment the physical units $G=c=1$. The field equations, in the metric formalism, for action \eqref{action} are obtained by the variational principle
\begin{equation}\label{field}
f_{\cal R}G_{\mu\nu}-\frac{1}{2}\left[f-f_{\cal R}{\cal R}\right]g_{\mu \nu }
-(\nabla_{\mu }\nabla _{\nu }-g_{\mu \nu }\Box )f_{\cal R}=8 \pi T_{\mu \nu }.
\end{equation}
Here $G_{\mu\nu}={\cal R}_{\mu\nu}-\frac{1}{2}{\cal R}g_{\mu\nu}$ is the
Einstein tensor, $f=f({\cal R})$, $f_{\cal R}({\cal R})=f_{\cal R}=df({\cal R})/d{\cal R}$ is the derivative of $f({\cal R})$ with
respect to the Ricci scalar and $T_{\mu \nu }$ is the
energy-momentum tensor of matter.

Spherically-symmetric solutions can be looked for, computing a point-like Lagrangian in which the spherically symmetry is placed in the action \eqref{action}. It is worth noting that a given symmetry can be imposed whether in the Lagrangian formalism, from which the Euler-Lagrange equations are subsequently derived, or directly into the field equations. The results are entirely equivalent. We will adopt the first strategy in order to define the space configuration  where the Noether vector acts on the point-like Lagrangian.
 
 A generic spherically-symmetric metric is:
\begin{equation}\label{me2}
{ds}^2=-A(r){dt}^2+B(r){dr}^2+C(r)d\Omega\,,
\end{equation}
where ${d\Omega}={d\theta}^2+{\sin\theta}^2{d\phi}^2$ is the
angular element. 
Imposing \eqref{me2} in the action 
(\ref{action}), in principle, a canonical form with a finite number of degrees of freedom may be assumed, that is 
 \begin{equation}\label{canonical}
\mathcal{A}=\int dr\mathcal{L}(A, A',  B, B', C, C', {\cal R}, {\cal R}')\,,
\end{equation}
where the Ricci scalar ${\cal R}$ and the metric coefficients $A$, $B$, $C$ are
the set of independent variables defining the space configuration  (see also \cite{arturo} for details). The prime indicates the derivative with respect to the radial
coordinate $r$. 

In order to obtain the point-like Lagrangian in the above coordinates, we write the action as
\begin{equation}\label{lm}
\mathcal{A}=\int
d^4x\sqrt{-g}\biggl[f-\lambda({\cal R}-\bar{\cal R})\biggr]\,,
\end{equation}
where $\lambda$ is a Lagrangian multiplier and $\bar{\cal R}$ is the
Ricci scalar expressed in terms of the metric (\ref{me2}),
i.e.~in more compact form, as
\begin{equation}
\bar{\cal R}={\cal R}^*+\frac{A''}{AB}+2\frac{C''}{BC}\,,
\end{equation}
where ${\cal R}^*$ collects first order derivative terms
\begin{eqnarray}
{\cal R}^*=\frac{A'C'}{ABC}-\frac{A'^2}{2A^2B}-\frac{C'^2}{2BC^2}-\frac{A'B'}{2AB^2}-\frac{B'C'}
{B^2C}-\frac{2}{C}\,.\nonumber\\
\end{eqnarray}
Varying  the action (\ref{lm})
with respect to ${\cal R}$ we obtain that $\lambda= f_{\cal R}$.
 Then, the action \eqref{action} becomes
\begin{eqnarray}\label{ac1}
{\cal A}&&=\int dr C\sqrt{A}\sqrt{B}\biggl[f-f_{\cal R}\biggl({\cal R}-{\cal R}^*-\frac{A''}{AB}-2\frac{C''}{BC}\biggr)\biggr]\nonumber\\&&
=\int dr 
\biggl\{C \sqrt{A}\sqrt{B}\biggl[f-f_{\cal R}({\cal R}-{\cal R}^*)\biggr] \nonumber\\&&\quad-{\,}\frac{f_{\cal {R}}C' A'}{(\sqrt{A})'(\sqrt{B})'}-
2\frac{(\sqrt{A})'}{(\sqrt{B})'}f_{\cal R}
C'\biggr\}
\nonumber\,.
\end{eqnarray}
Then the canonical point-like Lagrangian is
\begin{eqnarray}\label{lag}
\mathcal{L}&&=-\frac{\sqrt{A}f_{\cal R}}{2C\sqrt{B}}{C'}^2-\frac{f_{\cal R}}{\sqrt{A B}}A'C'-\frac{Cf_{\cal RR
}}{\sqrt{A B}}A'{\cal R}'+\nonumber\\\nonumber\\&&\quad-\frac{2\sqrt{A}f_{\cal RR}}{\sqrt{B}}{\cal R}'C'-\sqrt{A B}[(2+C{\cal R})f_{\cal R}-Cf]\,.\nonumber\\
\end{eqnarray}
The above Lagrangian can be recast in a suitable form
 introducing the
matrix formalism:
\begin{equation}\label{la}
\mathcal{L}={\underline{q}'}^{\rm T}\hat{T}\underline{q}'+V\,,
\end{equation}
where $\underline{q}=(A,B,C,R)$ and $\underline{q}'=(A',B',C',R')$
are the generalized positions and velocities associated with
$\mathcal{L}$. The index ${\rm T}$ indicates the transposed column
vector. The kinetic tensor is given by ${\displaystyle
\hat{T}_{ij}\,=\,\frac{\partial^2\mathcal{L}}{\partial q'_i\partial
q'_j}}$.  $V=V(q)$ is the potential depending only on the
configuration variables.

The general form of the Euler\,-\,Lagrange equations is
\begin{eqnarray}\label{fe2}\nonumber
\frac{d}{dr}\nabla_{q'}\mathcal{L}-\nabla_{q}\mathcal{L}=2\frac{d}{dr}\biggl(\hat{T}\underline{q}'\biggr)-\nabla_{q}V-{\underline
{q}'}^{\rm T}\biggl(\nabla_{q}\hat{T}\biggr)\underline{q}'\,=\nonumber\\\nonumber\\=\,2\hat{T}\underline{q}''+2\biggl(\underline{q}'\cdot\nabla_{q}\hat{T}\biggr)
\underline{q}'-\nabla_{q}V-\underline{q}'^{\rm T}\biggl(\nabla_{q}\hat{T}\biggr)\underline{q}'=0\,,\nonumber\\
\end{eqnarray}
which gives the equations of motion in terms of  $A$, $B$, $C$
and ${\cal R}$, respectively. After some manipulations, it is possible to demonstrate that the variable $B$ can be expressed as a combination of $A$ and $C$, that is 
\begin{eqnarray}\label{eqb}
B&=&\left(2C^2f_{\cal RR}A'{\cal R}'+2Cf_{\cal R}A'C'+4ACf_{\cal RR}M'{\cal R}'
\right.\nonumber\\&& \left.+Af_{\cal R}C'^2\right) \times \left(2AC[(2+C{\cal R})f_{\cal R}-Cf]\right)^{-1}\,.\nonumber\\
\end{eqnarray}
By inserting Eq.~(\ref{eqb}) into the Lagrangian (\ref{lag}), we
obtain a non-vanishing Hessian matrix which removes the singular
dynamics, and then the Lagrangian \eqref{lag} may be recast in the more manageable form
\begin{eqnarray}\label{lag2}
{\bf L}&&=\frac{[(2+C{\cal R})f_{\cal R}-fC]}{C}
[2C^2f_{\cal RR}A'{\cal R}'\nonumber\\&&\quad
+2CC'(f_{\cal R}A'+2Af_{\cal RR}{\cal R}') +Af_{\cal R}C'^2]\,. 
\end{eqnarray}
Since ${\displaystyle \frac{\partial{\bf L}}{\partial r}=0}$,
${\bf L}$ is canonical (${\bf L}$ is the quadratic form of
generalized velocities, $A'$, $C'$ and $R'$ and then coincides
with the Hamiltonian), so that we can consider ${\bf L}$ as a
 Lagrangian with three  degrees of freedom. 
\section{Spherically symmetric solutions via Noether symmetry approach}
\label{III}
We now search for symmetries for the Lagrangian (\ref{lag2}) in order to obtain exact solutions. 
It is known that if the following relation holds
\begin{eqnarray}\label{NS}
L_X{\bf L}=0\,, \rightarrow X{\bf L}=0\,,
\end{eqnarray}
then Noether symmetries exist. Here $L_X$ is the Lie derivative with respect to the Noether vector
\begin{equation}
\label{vectorN}
X\equiv\underline{\alpha}\nabla_q+\underline{\alpha}'\nabla_{q'}\,,
\end{equation}
$\underline{\alpha}$ are functions of configuration variables and $\underline{\alpha} '$ their  derivatives.
The second part of Equation \eqref{NS} means that the vector derivative $X$ is applied to the Lagrangian ${\bf L}$. Being, for example,  ${\displaystyle X=\alpha\frac{\partial}{\partial q_i}+\dot{\alpha}\frac{\partial}{\partial \dot{q}_i}}$,  it is  ${\displaystyle X{\bf L}=\alpha\frac{\partial {\bf L}}{\partial q_i}+\dot{\alpha}\frac{\partial {\bf L}}{\partial \dot{q}_i}}$, That is the contraction of ${\bf X}$ on ${\bf L}$.

In general, Equation \eqref{NS} is the contraction of the Noether vector $X$ on the tangent space  
 $\mathcal{TQ}=\{A, A', C, C', {\cal R}, {\cal R}'\}$ with the space of the configuration 
given by $\mathcal{Q}=\{A, C, {\cal R}\}$.
Explicitly, we have:
\begin{equation}\label{NS1}
L_{\mathbf{X}}{\bf L}\,=\,\underline{\alpha}\cdot\nabla_{q}{\bf L
}+\underline{\alpha}'\cdot\nabla_{q'}{\bf L}
=\underline{q}'^{\rm T}\biggl[\underline{\alpha}\cdot\nabla_{q}\hat{{\bf
L }}+ 2\biggl(\nabla_{q}\alpha\biggr)^{\rm T}\hat{{\bf L
}}\biggr]\underline{q}'\,,
\end{equation}
where, in the matrix formalism, it is
${\bf L}=\underline{q'}^{\rm T}\hat{{\bf
L}}\underline{q'}$.
Equation \eqref{NS1} vanishes if the functions ${\underline{\alpha}}$ satisfy the
following system
\begin{equation}\label{sys}
\underline{\alpha}\cdot\nabla_{q}\hat{{\bf L}}
+2(\nabla_{q}{\underline{\alpha}})^{\rm T}\hat{{\bf L
}}\,=\,0\,\longrightarrow\ \ \alpha_{i}\frac{\partial
\hat{{\bf L}}_{km}}{\partial
q_{i}}+2\frac{\partial\alpha_{i}}{\partial q_{k}}\hat{{\bf L
}}_{im}=0\,.
\end{equation}
The functions
$\alpha_{i}$, which fix the Noether vector, are obtained by solving the system \eqref{sys}.
The system of equations \eqref{sys} is related to the form of $f({\cal R})$-Lagrangian. In particular, classes of $f({\cal R})$ models, consistent with the spherical symmetry, are determined by solving the above system \cite{arturo}.
Conversely, by choosing the $f({\cal R})$ form, we can explicitly solve (\ref{sys}). 
We find that the system (\ref{sys}) is satisfied for
\begin{equation}\label{solsy}
f({\cal R})\,=\,f_0 {\cal R}^{1+\epsilon}\,,\end{equation}
and 
\begin{equation}
\underline{\alpha}=(\alpha_1,\alpha_2,\alpha_3)=
\biggl[(1-2\epsilon)kA,\ -kC,\ k{\cal R}\biggr]\,,
\end{equation}
where $\epsilon$ is any real number, $k$ an integration constant and $f_0$ a
dimensional coupling constant. Eq.\eqref{solsy} is not the unique possible $f(\cal R) $ solution that can be derived from the  Noether Symmetry Approach, however it is the only analytic and available in explicit form \cite{libroSV}.
This means that for any $f({\cal R})=f_0{\cal R}^{1+\epsilon}$, a Noether
symmetry exists and it is related to a constant of motion $\Sigma_{0}$ given by the equations of motion, that is 
\begin{eqnarray}\label{cm}\nonumber
\Sigma_{0}\,&&=\,\underline{\alpha}\cdot\nabla_{q'}{\bf L}
=
2(1+\epsilon)kC{\cal R}^{2\epsilon-1}[2(1+\epsilon)+\epsilon C{\cal R}]\times\nonumber\\\nonumber\\&&\quad\times[(\epsilon-1){\cal R}A'-(2(1+\epsilon)^2-3\epsilon-2)A{\cal R}']\,.\end{eqnarray}
A physical interpretation of $\Sigma_{0}$ is possible by starting from General Relativity, i.e. $\epsilon =0$. 
In this case, the Noether symmetry yields the solution
\begin{equation}\label{solsygr}\underline{\alpha}_{GR}=(-kA,
kC)\,,\quad f({\cal R})=f_0 {\cal R}\,.
\end{equation}
The functions $A$ and $C$ give the Schwarzschild solution 
and then, upon restoration of standard units, the constant of motion is
\begin{equation}
\label{cmgr}\Sigma_{0}= \frac{2GM}{c^2}\,,
\end{equation}
where $M$ is the gravitational mass of the system. In other words, in the case of Einstein gravity, the Noether
symmetry gives    the Schwarzschild radius (and the gravitational mass) as a conserved quantity.
In the  general case  \eqref{solsy}, the Lagrangian (\ref{lag2})
becomes, 
\begin{eqnarray}\label{89}{\bf L}&=&\frac{(1+\epsilon){\cal R}^{2\epsilon-1}[2(1+\epsilon)+\epsilon C{\cal R}]}{C}
\left[2\epsilon C^2A'{\cal R}'+\right.\nonumber\\\nonumber\\&&\left.+2C{\cal R}C'A'+4\epsilon ACC'{\cal R}'+A{\cal R}C'^2\right]\,,\end{eqnarray}
and exact solutions, using the constant of motion,  can be given in the form 
\begin{eqnarray}\label{B}B&&=\frac{1+\epsilon}{2AC{\cal R}[2(1+\epsilon)+\epsilon C{\cal R}]}\left[2\epsilon C^2A'{\cal R}'+\right.\nonumber\\&&\quad\left.+2C{\cal R}C'A'+4\epsilon ACC'{\cal R}'+A{\cal R}C'^2\right]\,,\end{eqnarray}

\begin{eqnarray}\label{}A={\cal R}^{\frac{\epsilon(2\epsilon+1)}{\epsilon-1}}\biggl\{k_1+\Sigma_{0}\int\frac{{\cal R}^{\frac{\epsilon(4\epsilon-1)}{1-\epsilon}}dr}{2k(\epsilon^2-1)C[2(\epsilon+1)+\epsilon C{\cal R}]}\biggr\}\nonumber\\\end{eqnarray}
where $k_1$ an integration constant. 
General Relativity is clearly recovered for $\epsilon =0$.
Such solutions can be used to obtain TOV solutions and $M-R$ relations parameterized by $\epsilon$.
Reversing the problem, the $M-R$ relation fixes the underlying theory of gravity, corrected with respect to General Relativity.
\section{Noether's stars}
\label{IV}
The above relations enable general solutions for the field
equations to be determined, giving the dependence of the scalar curvature $\cal R$ vs the radial
coordinate $r$.
The first step is to calculate the 
interior metric solution that must be matched with the corresponding exterior solution. 
In order to restore the TOV standard notation, let us set
$A(r)=e^{2\psi}$, $B(r)=e^{2\lambda}$, $C(r)=r^2$, where $\psi$ and $\lambda$ are functions of the radial coordinate $r$ only \footnote{$C(r)$ is the function that rules how $2D$ surfaces, embedded in spacetime,  are measured. Choosing $C(r)=r^2$ implies
that the length of a circle, centered in the origin of the coordinates, is
$2\pi r$ (i.e. in such a way we preserve the spherical symmetry). If $C(r)\neq r^2$, the circle is deformed. Furthermore, the system can present singularities if $C(r)$ is not continuos and derivable.  These cases can be  interesting in the cases of anisotropic and/or inhomogeneous collapses.  }. 
The metric (\ref{me2}) can then be recast in the standard form:
\begin{equation}\label{metric}
    ds^2= -e^{2\psi}dt^2 +e^{2\lambda}dr^2 +r^2 d\Omega^2\,. 
\end{equation}
The energy-momentum tensor is 
\begin{equation}
 T_{\mu\nu}=\mbox{diag}\left(e^{2\psi}\rho, e^{2\lambda}p, r^2p,
 r^{2}\sin^{2}\theta\,p \right)\,,\end{equation}
 where $\rho$ is the matter density and $p$ is the pressure \cite{weinberg}. The nontrivial components of the  field equations \eqref{field} give the TOV equations for $f({\cal R})$ gravity \cite{asta}, which in our case, for  $f({\cal R})=f_0{\cal R}^{1+\epsilon}$,  are:

\begin{eqnarray} \label{TOV1}&&\frac{{\cal R}^{\epsilon}}{r^2}\left[r\left(1-e^{-2\lambda }\right)\right]=8\pi
\rho+\frac{1}{2} \epsilon {\cal R}^{1+\epsilon}+
\nonumber\\&&+
e^{-2\lambda}\bigg\{\left(\frac{2}{r}-\frac{d\lambda}{dr}\right)\left[\epsilon(1+\epsilon){\cal R}^{\epsilon-1} {\cal R}'\right]
+\nonumber\\&&
+\biggl[\epsilon (1+\epsilon) {\cal R}^{\epsilon-2} \left[{\cal R} {\cal R}''+(\epsilon-1) {\cal R}'^2\right)\biggr]\biggr\}\,,
 \end{eqnarray}

\begin{eqnarray}\label{TOV2}&& \frac{{\cal R}^{\epsilon}}{r}\left[2e^{-2\lambda}\frac{d\psi}{dr}-\frac{1}{r}\left(1-e^{-2\lambda}\right)\right]
=8\pi
p+\nonumber\\&&+\frac{1}{2}\epsilon {\cal R}^{1+\epsilon}
+e^{-2\lambda}\left(\frac{2}{r}+\frac{d\psi}{dr}\right)\left[\epsilon (1+\epsilon) {\cal R}^{\epsilon-1} {\cal R}'\right]\,.\nonumber\\
 \end{eqnarray}
Here, now the prime indicate the derivative with respect $\cal R$. 
Adopting physical units, we may set  $f_0=1$.
For $\epsilon=0$, the standard TOV equations of General Relativity are recovered. 
The stellar configuration is a solution of the field equations and the conservation equations for the energy-momentum tensor, $\nabla^\mu
T_{\mu\nu}=0$, from which the hydrostatic equilibrium condition follows:
\begin{equation}\label{hydro}
    \frac{dp}{dr}=-(\rho
    +p)\frac{d\psi}{dr}.
\end{equation}
In $f(\cal{R})$
gravity, the scalar curvature is a dynamical variable and the equation
for ${\cal R}$ can be obtained by taking into account the trace of the field equations
(\ref{field}). 
We have
\begin{equation}\label{TOV3}
3\square f_{\cal R}+{\cal R}f_{\cal R}-2f=-{8\pi}(\rho-3p)\,,
\end{equation}
that explicitly becomes 
\begin{eqnarray}\label{TOV3tra}
&&(\epsilon-1) {\cal R}^{\epsilon+1}+3\epsilon(1+\epsilon) e^{2 \lambda}\biggl[(\epsilon-1) {\cal R}^{\epsilon-2} {\cal R}'^2+\nonumber\\&&+ {\cal R}^{\epsilon-1} {\cal R}' \left( \frac{2}{r}-\frac{d\lambda}{dr}+\frac{d\psi}{dr}\right)+ {\cal R}^{\epsilon-1} {\cal R}''\biggr]=\nonumber\\&&=-{8\pi}(\rho-3p)\,.
\end{eqnarray}
The above equation give us a further constraint to solve the TOV equations \cite{asta}. 
These equations \eqref{TOV1}--\eqref{TOV3} can be solved by numerical integration from $r = 0$, but we require a set of boundary
conditions to fix the integration constants, and an equation of state that gives a
relation between the density and pressure (see e.g.~\cite{MRnostro} for details on the numerical method).

\begin{figure}
\includegraphics[angle=-90,scale=0.35]{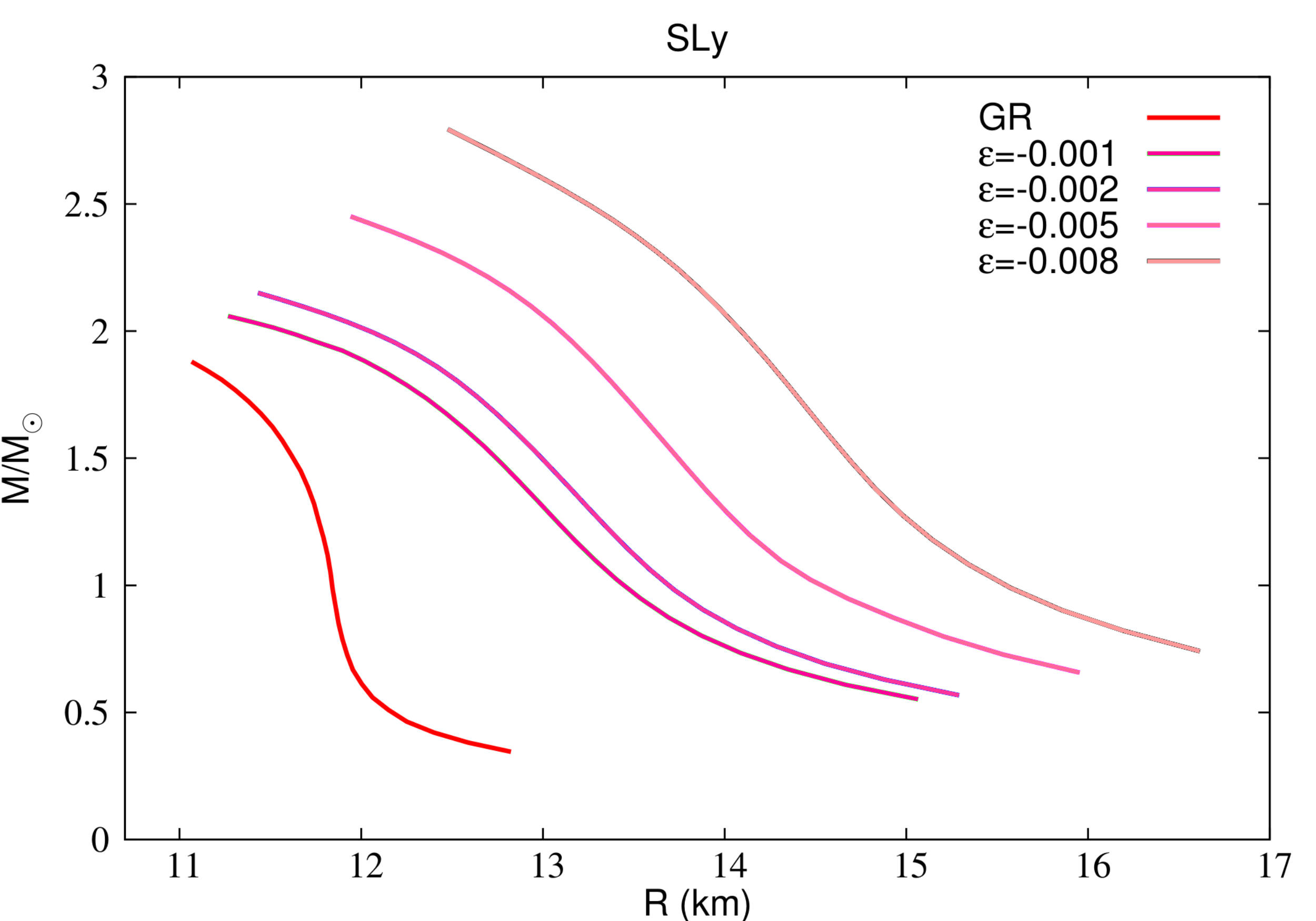}
\caption{$M-R$ diagram for  $f({\cal R})={\cal R}^{1+\epsilon}$ for the Sly EoS with different values of $\epsilon$ (purple color scales). The classical TOV solution corresponding to $\epsilon=0$ is also shown as a red line.}
 \label{MRsly}
\end{figure}

\begin{figure}
\includegraphics[angle=-90,scale=0.34]{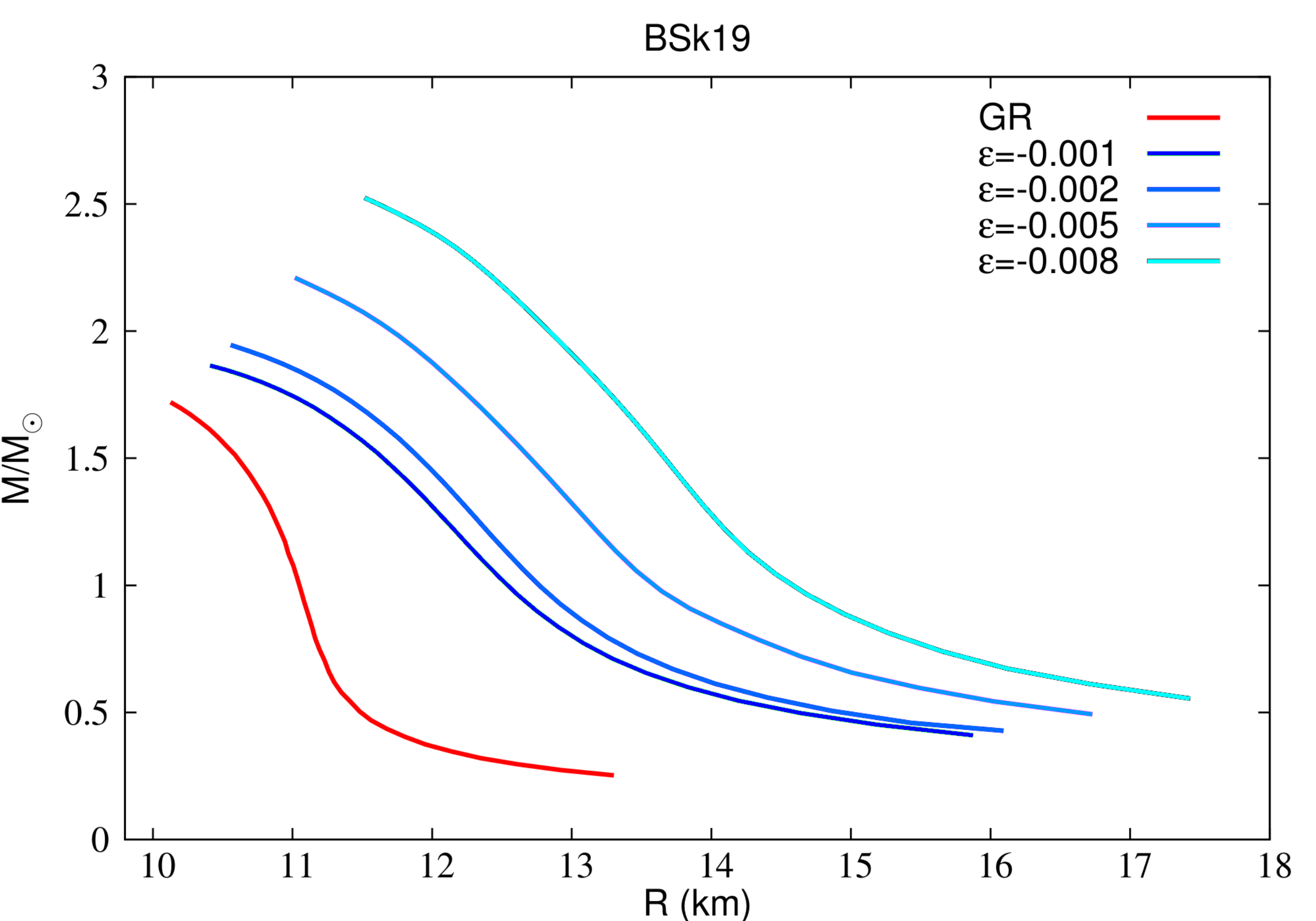}
\caption{$M-R$ diagrams for (\ref{solsy}) using the Bsk19 EoS with different values of $\epsilon$ represented as a blue color scale. The classical TOV solution corresponding to $\epsilon =0$ is also shown as a red line.}
 \label{MRbsk19}
\end{figure}

\begin{figure}
\includegraphics[angle=-90,scale=0.34]{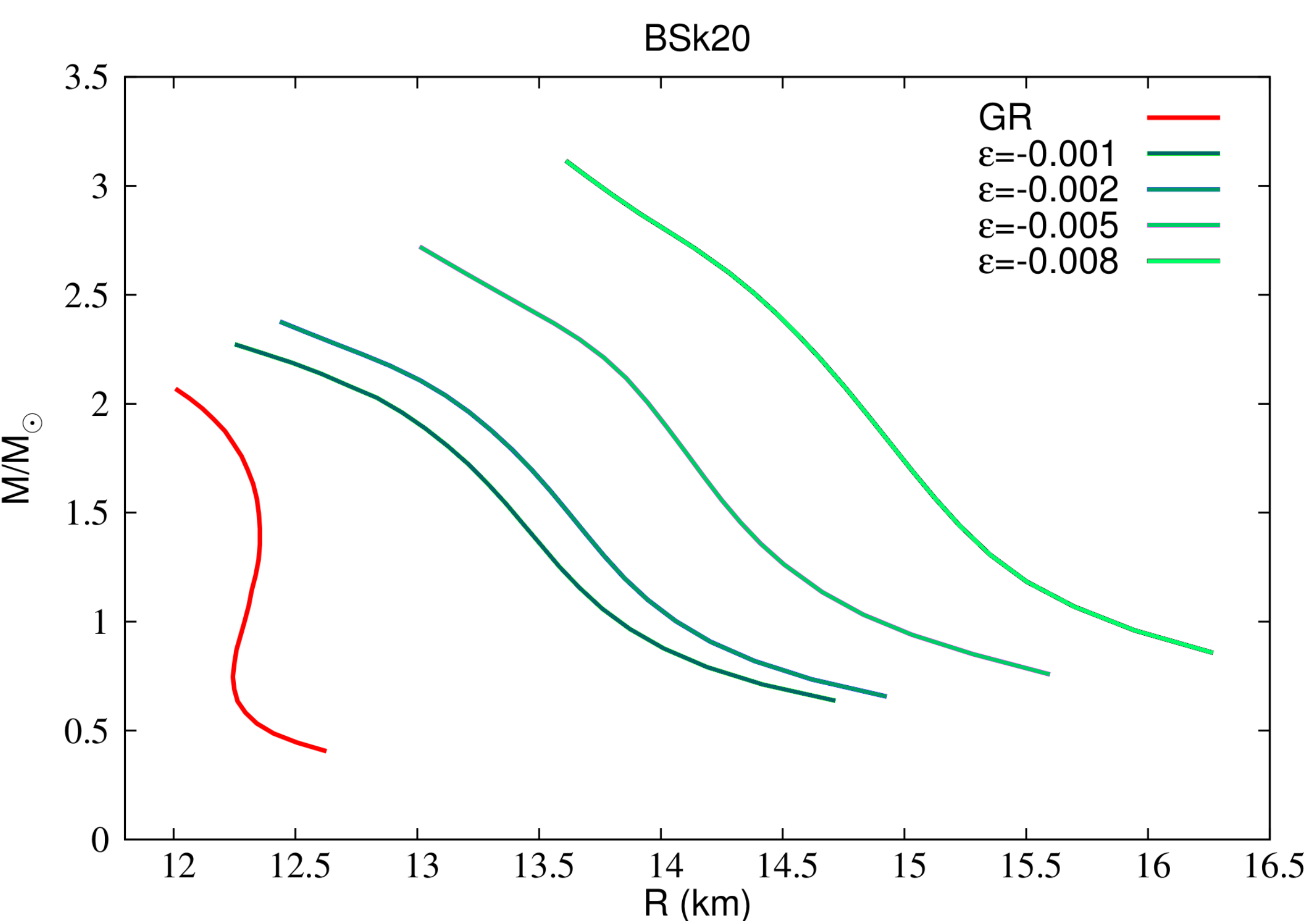}
\caption{$M-R$ diagrams for \eqref{solsy} using EoS  BSk20. The green color scale represent the different values of $\epsilon$. The classical TOV corresponding to $\epsilon =0$ is also shown as a red line.}
 \label{MRbsk20}
\end{figure}

\begin{figure}
\includegraphics[angle=-90,scale=0.34]{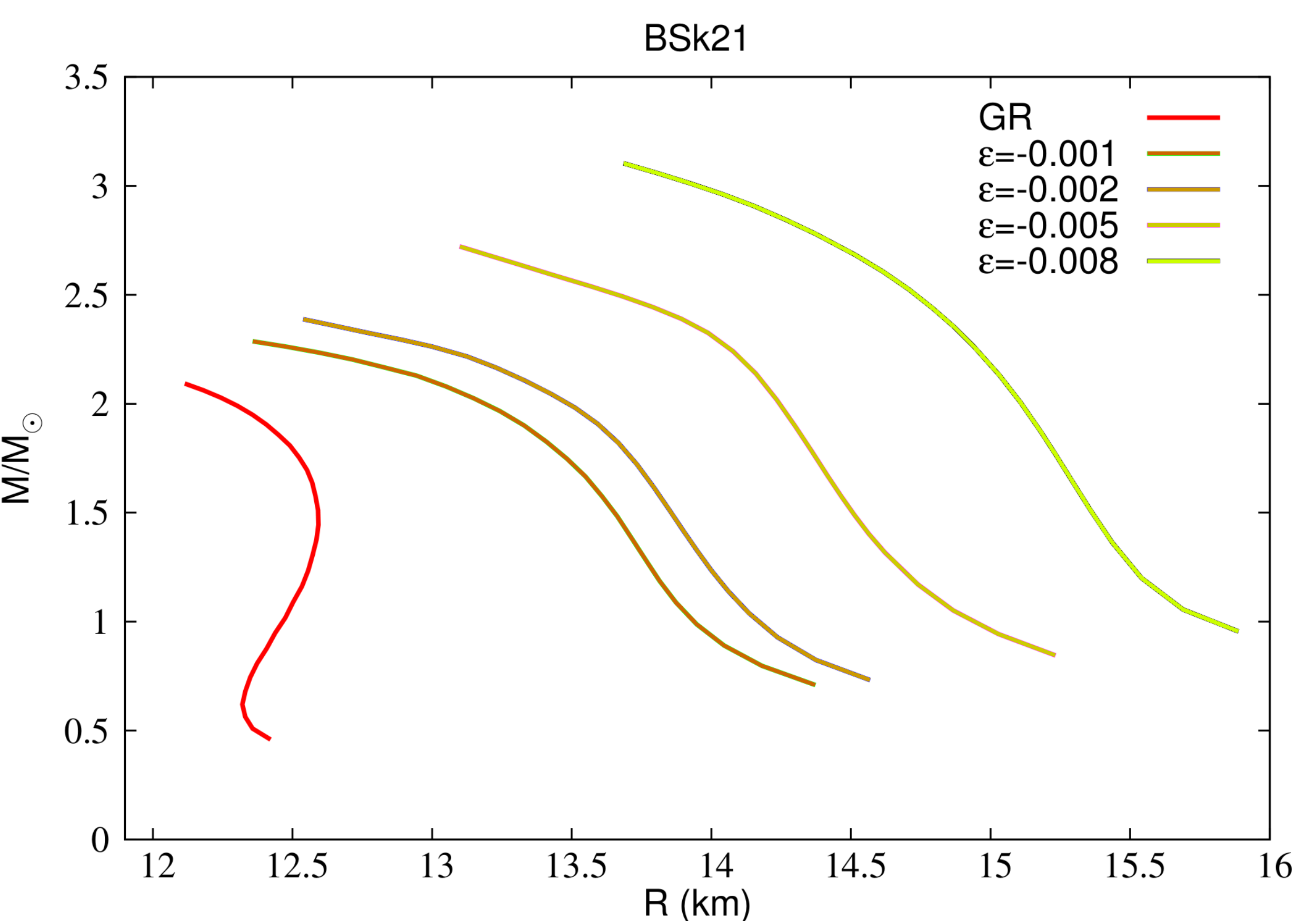}
\caption{$M-R$ diagrams for  $f({\cal R})$ given in equation (\ref{solsy}). Here different curves for different values of $\epsilon$ (yellow color scale) using the BSk21 EoS are shown. The classical TOV corresponding to $\epsilon =0$ is also shown as a red line.}
 \label{MRbsk21}
\end{figure}

In Figs.~\ref{MRsly}--\ref{MRbsk21},  the $M-R$ diagram for various values of $\epsilon$ is represented. 
Herein some popular equations of state are used, namely Sly, BSK19, BSK20 and BSK21 respectively \cite{hp2004,potekhin13}. 
It is clear to see that  for $|\epsilon| > 0.01$ there is a significant deviation with respect to General Relativity. 
Noteworthy is the fact that, for increasingly large values of $|\epsilon|$, the $M-R$ diagrams assume a self-similar behaviour. 
Larger radii and masses are achieved for negative values of the scaling parameter while, in the case of positive values, the traces are bent with usual General Relativity TOV equations. It is straightforward to see that we can reach masses about $(2.8-3)M_{\odot}$ using the BSK20 and BSK21 equations of state. 

 A final comment is in order at this point. The radius in the figures has not to be identified with the constant of motion. The constant of motion fixes the functional relation between the mass M and the radius R saying that there is a characteristic gravitational radius which coincides with  the Schwarzschild  radius of General Relativity, i.e. for $\epsilon =0$. Clearly, for any $\epsilon$ the gravitational  radius changes.  The integration constant $k$ can be chosen equal to 1 without affecting the system. The sign of  $\epsilon$ is related to   the $(M-R)$ relation.  If $\epsilon <0$  larger stars can be achieved with respect to General Relativity. For $\epsilon >0$, we obtains  smaller stars.

\section{Conclusions} 
\label{V}
The mass of a self-gravitating system can be considered as a Noether charge according to the existence of the Noether symmetries. 
In this paper, we derived both the conserved quantities and the functional form of $f({\cal R})$ gravity according to the so-called Noether Symmetry Approach \cite{cimento}. 
The final output is that a power-law form of $f({\cal R})$ gravity is determined by the Noether vector. 
The power $\epsilon$ can be any real number. 
Such a parameter is useful in order to study deviations with respect to General Relativity.\\
In particular, spherically-symmetric solutions are considered and we  derived the field equations parameterized by $\epsilon$.
Starting from this scheme, modified TOV equations are obtained and, assuming reliable equations of state discussed in the literature, the $M-R$ relation is achieved. 
According to the value and the sign of $\epsilon$, it is possible to  show that radii and masses of compact neutron stars change with respect to General Relativity. This fact  allows, in principle,  that  larger/smaller objects can be obtained by varying the gravitational sector with respect to those provided by the standard theory.
In particular, extremely large objects could be framed depending on modified gravity \cite{asta}. \\
Some considerations are in order at this point.
The first is related to  the Noether symmetries. The associated conserved quantity leads the $M-R$ relation. In other words, the existence of the symmetry is capable of ruling the stellar parameters and then the position of the star on the Hertzsprung-Russell diagram. In a general sense, the whole diagram could depend on the given theory of gravity and compact objects, where strong field effects are effective, could be 
 a useful testbed to retain or rule out alternative models.\\
 Another consideration is related to the role of gravity in this framework. It seems that the parameter $\epsilon$ 	can really point out  deviations with respect General Relativity   emerging  at given interaction lengths. Such lengths, depending on $\epsilon$,  has a similar role of the Schwarzschild radius (derived for $\epsilon =0$). The paradigm is that any theory of gravity has its own characteristic gravitational radius that can be something else with respect to the standard one of General Relativity. It is worth noticing that for small deviation with respect to General Relativity we can write
 \begin{equation}{\cal R}^{1+\epsilon}\simeq {\cal R}+\epsilon {\cal R} \ln {\cal R}+{\cal O}(\epsilon^2)\,,
 \end{equation}
 and then control the magnitude of the corrections with respect to the standard Hilbert-Einstein action. Such deviation could come out in the strong field regimes inside compact objects that could be very similar to some situations present in the early universe where logarithmic corrections emerge from quantization of curved space time\cite{asta,Starobinsky80}. 
 
 Finally, neutron stars, achieved in such a framework, could really discriminate between modified gravity  and dark matter scenarios: in fact no exotic particle is requested in this context. The only natural assumption is that a symmetry breaking of gravitational interaction can happen at a given  scale and  energy, exactly like in the case of Starobinsky model of early universe where higher order curvature terms like ${\cal R}^2$ give rise to  inflation \cite{libroSV,Starobinsky80}.
 
The Noether Symmetry Approach deserves some further general considerations. As firstly discussed in \cite{cimento}, the utility of the method is twofold. From one hand, it allows to find out exact solutions since the presence of Noether symmetries reduces the related dynamical systems. Clearly, if the number of symmetries coincides with the number of dimensions of configuration space, the system is completely integrable. On the other hand, as shown here, the approach allows to select the class  of models, in this case the power-law form of $f(\cal R)$ gravity. This means that the further degrees of freedom of any modified theory of gravity (scalar tensor, vector tensor, and so on) can be linked to the symmetries that rule the dynamics (see \cite{cimento} for scalar tensor gravity). In this perspective, the Noether Symmetry Approach is a general criterion  to select viable theories of gravity.
 
\section*{Acknowledgments}
The Author is supported by the Grant "BlackHoleCam" Imaging the Event Horizon of Black Holes awarded by the ERC in 2013 (Grant No. 610058). The Author acknowledges the COST Action CA15117 (CANTATA) and INFN Sez. di Napoli (Iniziative Specifiche QGSKY and TEONGRAV).



\end{document}